\documentclass{article}
\usepackage[utf8]{inputenc}
\usepackage{graphicx}
\usepackage{subcaption}
\usepackage{color}
\usepackage{url}
\usepackage{amsmath}

\title{Molecular simulations matching denaturation experiments for N6-methyladenosine}
\author{Valerio Piomponi, Thorben Fr\"ohlking, Mattia Bernetti, and Giovanni Bussi}
\date{
Scuola Internazionale Superiore di Studi Avanzati, SISSA, \\via Bonomea 265, 34136 Trieste, Italy}

\begin{document}

\maketitle

\begin{abstract}
Post-transcriptional modifications are crucial for RNA function and can affect its structure and dynamics.
Force-field based classical molecular dynamics simulations are a fundamental tool to characterize biomolecular dynamics
and their application to RNA is flourishing.
Here we show that the set of force-field parameters for N$^6$-methyladenosine (m$^6$A) developed for the commonly used AMBER force field
does not reproduce duplex denaturation experiments and, specifically,
cannot be used to describe both paired and unpaired states.
Then we use reweighting techniques to derive new parameters  matching available experimental data.
The resulting force field can be used to properly describe paired and unpaired m$^6$A in both \emph{syn} and \emph{anti} conformation,
and thus opens the way to the use of molecular simulations to investigate the effects of N6 methylations on RNA structural dynamics.
\end{abstract}

\section{Introduction}
\begin{figure}
\begin{center}
\includegraphics[width=0.5\textwidth]{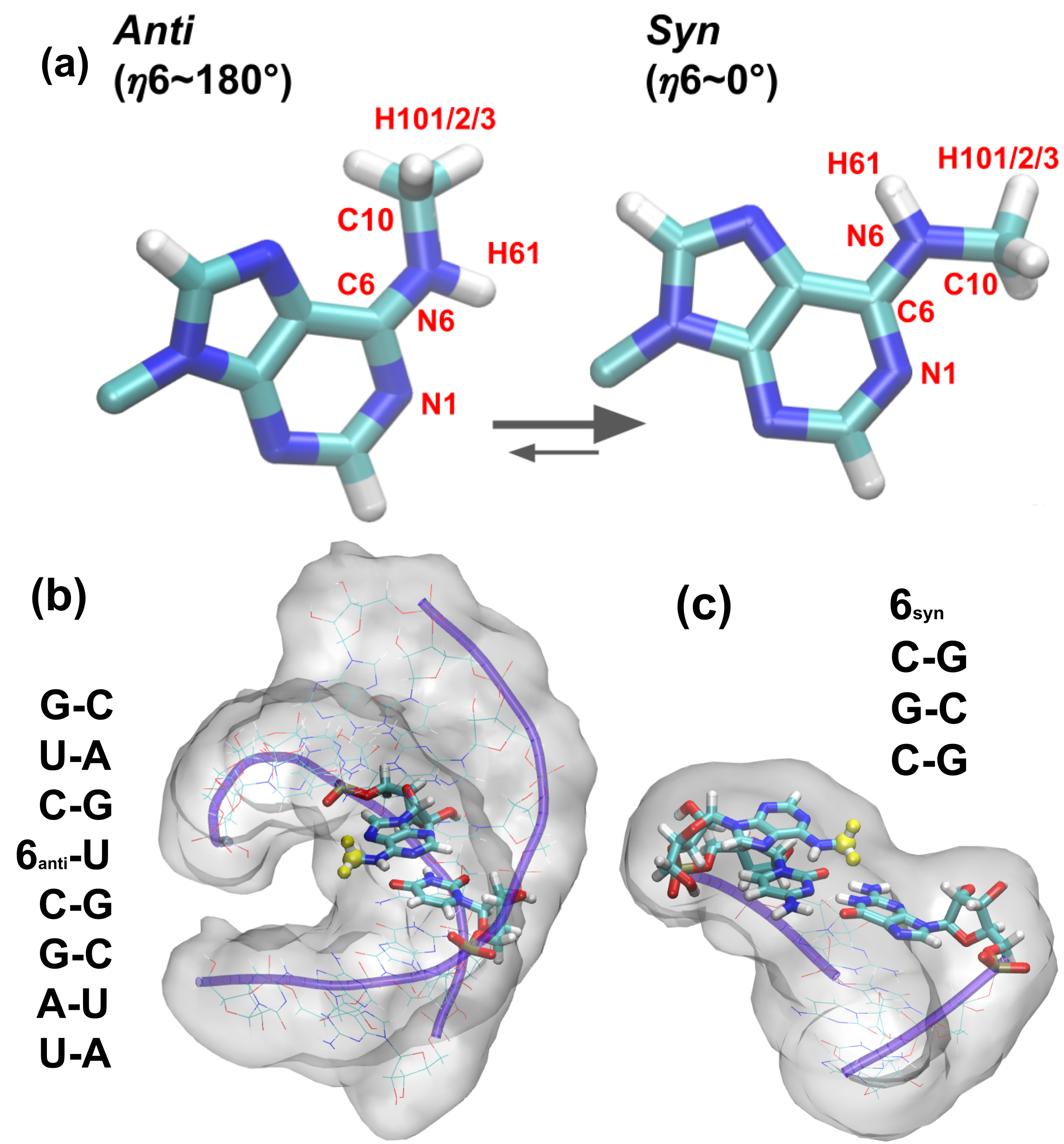}
\end{center}
\caption{N$^6$-methyladenosine (m$^6$A) nucleobase in \emph{anti} (less stable) and \emph{syn} (more stable) conformations (panel a) \cite{roost2015structure, liu2020quantitative}.
Atom names in red correspond to charges reparametrized in this work.
Example of a duplex containing m$^6$A in \emph{anti} conformation, which is the expected conformation for the nucleotide when Watson--Crick paired (panel b).
A 6 is used to denote m$^6$A in secondary structures for compactness.
Example of a duplex with m$^6$A as a dangling end in \emph{syn} conformation (panel c).
The m$^6$A methyl group is highlighted in yellow.
}
\label{fig1}
\end{figure}

Post-transcriptionally modified nucleotides are crucial for RNA function \cite{gilbert2016messenger,harcourt2017chemical}.
Methylation of adenine in the N6 position (m$^6$A) is the most prevalent chemical modification in messenger RNAs
and has been observed both in coding and non-coding RNAs \cite{gilbert2016messenger,harcourt2017chemical,patil2016m,he2021m6a}.
m$^6$A can fine regulate the interaction of RNA with specific proteins known as m$^6$A readers.
In addition, similarly to other chemical modifications \cite{tanzer2019rna}, it can directly affect RNA
stability and structural dynamics (see, e.g., \cite{liu2015n,huang2017control,jones2022structural}).
Specifically, m$^6$A has been suggested to weaken Watson--Crick pairings
due to the incompatibility of its most stable conformation (\emph{syn}) with duplex formation \cite{roost2015structure,hopfinger2020predictions,liu2020quantitative} (see also Fig.~\ref{fig1}).
Interestingly, recent nuclear magnetic resonance (NMR) experiments have identified \emph{syn}/\emph{anti} dynamics in
both paired and unpaired m$^6$A, recapitulating the effect of N6 methylation on RNA conformational kinetics \cite{liu2020quantitative}.
Molecular dynamics (MD) simulations give access to structural dynamics at the atomistic resolution \cite{sponer2018rna}
and are thus an ideal tool to complement NMR studies.
The capability of classical force fields to predict the dynamics of difficult structural motifs is steadily increasing
\cite{sponer2018rna,tan2018rna,frohlking2022automatic}.
However, the number of applications of MD simulations to N6-methylated RNAs reported to date is still limited
\cite{liu2020quantitative,xu2016additive,li2019flexible,li2021atomistic,hurst2021deciphering,krepl2021recognition,purslow2021n}.
Force fields parameters for m$^6$A developed by Aduri \emph{et al.} \cite{aduri2007amber} 
were determined in a bottom-up fashion and
are compatible with the AMBER force field, which is widely adopted for RNA simulations \cite{sponer2018rna}.
These parameters are the default choice for a structure refinement tool \cite{stasiewicz2019qrnas}.
However, they have been quantitatively validated against a limited set of experimental results \cite{hurst2021deciphering}.
The recent publication of denaturation experiments on a number m$^6$A-containing duplexes \cite{roost2015structure,kierzek2022secondary}
calls for a more extensive validation of force-field parameters and, ideally, for fitting force-field parameters directly on
experiments \cite{frohlking2020toward}.

\begin{figure}
\begin{center}
\includegraphics[width=0.5\textwidth]{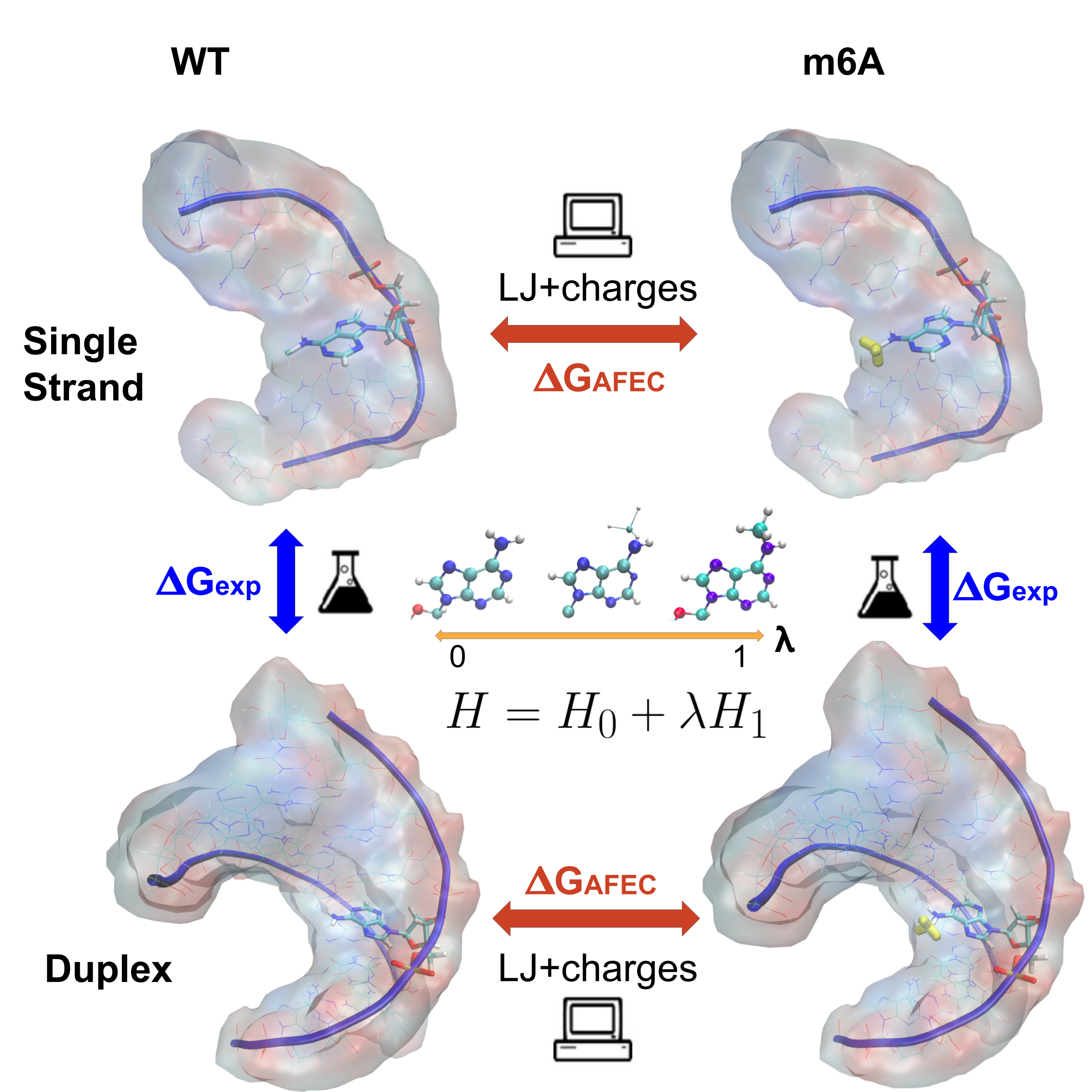}
\end{center}
\caption{Thermodynamic cycle. Alchemical free-energy calculations (AFEC) allow computing $\Delta G$ by integrating along an alchemical path $\lambda$ describing the transformation of non methylated adenosine into m$^6$A, by switching \emph{on/off} non-bonded interaction of specifically chosen atoms. The relative free-energy change due to the modification can be estimated as the $\Delta \Delta G$ between AFECs performed on a duplex and on the corresponding single strand. This quantity can be directly compared to the difference in thermodynamic stability of duplexes with or without the modification, that can be measured experimentally through denaturation experiments.}
\label{thermocycle}
\end{figure}

In this paper, we validate and improve the parameters introduced in Ref.~\cite{aduri2007amber} by using alchemical free-energy
calculations (AFECs) \cite{mey2020best}. To this end, an unmodified adenine is converted to a modified one
by switching \emph{on/off} non-bonded interactions of specifically chosen atoms,
in both single-stranded and double-stranded RNAs \cite{hurst2021deciphering} (see also Fig.~\ref{thermocycle}).
We then develop a reweighting technique that can be used to predict results corresponding to a different set of charges
without the need to perform new MD simulations.
Additionally, we extend a recently-introduced force-field fitting strategy 
\cite{cesari2019fitting}
to be usable in the context of alchemical simulations.
The introduced approach allows training six charges and a dihedral potential so as to quantitatively reproduce
methylation effects in denaturation experiments.
The resulting force field can be used to properly describe paired and unpaired m$^6$A in both \emph{syn} and \emph{anti} conformation.

\section{Material and Methods}

\subsection{Simulated systems}

We simulated the isolated m$^6$A nucleoside,
9 m$^6$A-methylated duplexes for which denaturation experiments are available in literature \cite{roost2015structure,kierzek2022secondary} (see Table \ref{table1}),
and the corresponding single stranded RNAs.
For the isolated m$^6$A nucleoside, we computed the $\Delta G_{syn/anti}$ by taking the difference in the
$\Delta G$s obtained with AFEC by methylating the adenosine in  \emph{syn} or \emph{anti} conformations.
We then chose a number of systems from Ref.~\cite{roost2015structure}.
For systems A4 and A5, where m$^6$A is present as dangling end and thus unpaired, we only performed AFEC corresponding
to the \emph{syn} conformation.
For the other systems, we performed AFEC in the expected \emph{anti} conformation. 
For the A2 and A3 systems we additionally performed AFEC in the unexpected
\emph{syn} conformation as a validation (population reported in Ref.~\cite{liu2020quantitative} is $\approx 1$\%).
In addition, we chose 5 more systems from Ref.~\cite{kierzek2022secondary}, with the following criterion:
they have a single methylation per-strand and the methylation occurs in an internal position of the duplexes.
For all these systems, we performed AFEC in the expected \emph{anti} conformation.
\begin{table}
\begin{center}
\resizebox{5.5cm}{!}{%
\begin{tabular}{|l|c|r|}
\hline
\multicolumn{1}{|c|}{\textbf{}}            &            \textbf{System}            &\multicolumn{1}{c|}{$\Delta \Delta G$ (kJ/mol)}\\
\hline
A1            &        m$^6$A $\Delta G_{syn/anti}$        &        6.3 \cite{roost2015structure}   \\
\hline
A2           &    UACG6CUG             &            1.7 $\pm$ 0.9 \cite{roost2015structure}\\
      &  AUGCUGAC   & \\
\hline
A3           &    CGAU6GGU             &            7.1 $\pm$ 0.9 \cite{roost2015structure}\\
      &  GCUAUCCA   & \\
\hline
A4           &    6CGC             &            -2.5 $\pm$ 1.2 \cite{roost2015structure}\\
      &  \ GCG   & \\
\hline
A5           &    GCG6             &            -1.7 $\pm$ 0.9 \cite{roost2015structure}\\
      &  CGC \    & \\
\hline
B1           &    GUC6CUG               &            2.5 $\pm$ 2.1 \cite{kierzek2022secondary}\\
      &  CAGUGAC   & \\
\hline
B2           &    ACU6UAGU               &            2.1 $\pm$ 1.3 \cite{kierzek2022secondary}\\
      &  UGAU6UCA   & \\
\hline
B3           &    AGUU6ACU               &            5.4 $\pm$ 1.3 \cite{kierzek2022secondary}\\
      &  UCA6UUGA   & \\
\hline
B4           &   CGGUGC6UCG                &            8.6 $\pm$ 0.8 \cite{kierzek2022secondary}\\
      &  GCU6CGUGGC   & \\
\hline
B5           &   ACUUA6GU               &            1.7 $\pm$ 1.0 \cite{kierzek2022secondary}\\
      &  UG6AUUCA   & \\
\hline
\end{tabular}%
}
\caption{List of systems involved in the fitting and relative experimental $\Delta \Delta G$. The first system is the single nucleotide and the experimental value corresponds to $\Delta G_{syn/anti}$. In the A2--A5 and B1--B5, the ``6'' in the double strand sequences is used to identify m$^6$A for compactness. $\Delta \Delta G$s for systems A2--A5 were measured by Roost \emph{et al.} \cite{roost2015structure}. In A4 and A5, the m$^6$A is positioned as a dangling end and has a stabilizing effect on the duplex. Experiments for systems B1--B5 were performed by Kierzek \emph{et al.} \cite{kierzek2022secondary}. In B2--B5 systems, the methylation occurs in both strands, however the $\Delta \Delta G$s  reported are to be intended ``per-methylation" .}
\label{table1}
\end{center}
\end{table}

Starting structures for MD simulations were built using 
the proto--Nucleic Acid Builder \cite{alenaizan2021proto}. Single strands were generated by deleting one of the chains from duplex structures.
All the MD simulations were performed using a modified version of GROMACS 2020.3 \cite{abraham2015gromacs} which
also implements the stochastic cell rescaling barostat \cite{bernetti2020pressure}. The AMBER force field was used for RNA \cite{cornell1995second,perez2007refinement,zgarbova2011refinement},
TIP3 model for water \cite{jorgensen1983comparison},
and Joung and Cheatham parameters for ions \cite{joung2008determination}. As a starting parametrization for m$^6$A, we used AMBER adenosine parameters combined with modrna08 \cite{aduri2007amber} charges for the nucleobase, adjusted so as to preserve the total charge of the nucleoside.
Details on the implementation of these parameters and on the initial tests are reported in Section S1.
Charges are given in Table S2.
We refer to this parametrization as Aduri force field.

\subsection{Alchemical Simulations}
For AFECs, we included a hybrid adenosine with double topology
in the force-field definition: the first topology corresponding to standard adenosine, and the second one corresponding to m$^6$A. We used 16 replicas in which Lennard-Jones parameters and partial charges were simultaneously interpolated.
In order to avoid singularities due to electrostatic interaction when the repulsive LJ potential is switched off \cite{mey2020best},
we used soft-core potentials as implemented in GROMACS  (sc-alpha=0.5, sc-sigma=0.3, and sc-power=1) \cite{beutler1994avoiding}.
Simulation boxes consist in rhombic dodecahedrons containing RNA, water, Na$^+$ and Cl$^-$ ions with an excess salt concentration of 0.1~M. For a subset of the systems, further simulations were performed for a salt concentration of 1~M.
The systems were energy minimized and subjected a multi-step equilibration procedure for each replica: 100 ps of thermalization to 300 K in the NVT ensemble was conducted through the stochastic dynamics integrator (i.e., Langevin dynamics) \cite{goga2012efficient}, and other 100 ps were run in the NPT ensemble simulations using the Parrinello--Rahman barostat \cite{parrinello1981polymorphic}.
In production runs, the stochastic dynamics integrator was used in combination with the stochastic cell rescaling barostat \cite{bernetti2020pressure} to keep pressure at 1 bar. Equations of motion were integrated with a time-step of 2 fs. Long-range
electrostatic interactions were handled by particle-mesh Ewald
\cite{darden1993particle}. During production, Hamiltonian replica exchange was used proposing exchanges every 200 fs. The set of $\lambda$ values defining replica's Hamiltonians were chosen in such a way to guarantee transition probabilities above 20\% and as homogeneous as possible (see Section S2),
ensuring a sufficient phase space overlap between replicas.
Each replica was simulated for $10$ ns, for a total of $16 \times 10 $ ns $ = 160$ ns for each system. To decrease numerical errors in energy recalculations, trajectories were saved in uncompressed format.
At the end of the production phase, the 16 independent ``demuxed'' (i.e., continuous) trajectories
were processed to recompute energies for each of the 16 Hamiltonian functions so as to compute $\Delta G$ via
binless weighted histogram analysis method (WHAM) \cite{souaille2001extension,shirts2008statistically,tan2012theory}.
Specifically, for each trajectory a weight $w$ was found for each snapshot $x$
that allows computing statistics for the unmodified adenine as a weighted average over the set of concatenated replicas.
These weights were then used to compute the $\Delta G$ associated to the methylation as:
\begin{equation}
\Delta G^{AFEC} = -k_BT \log\left[\frac{\sum_x w(x) e^{-\beta \Delta E(x)}}{\sum_x  w(x)}\right]
\label{eq:deltag}
\end{equation}
where $\Delta E (x) = E_{\lambda = 1}(x) -E_{\lambda=0}(x)$ is the difference between the total energy computed
with the Hamiltonian energy functions associated to m$^6$A and adenosine, respectively.
We used a bootstrapping procedure to compute the statistical error on $\Delta G$ estimates by resampling the 16 continuous trajectories 200 times with replacement \cite{efron1986bootstrap}. As a control, we computed $\Delta G$s using the standard Bennett-acceptance-rate estimate implemented in Gromacs \cite{bennett1976, wu2005bar}.

$\Delta \Delta G$s were obtained taking the difference between $\Delta G$s obtained by methylating the adenosine in \emph{anti} or \emph{syn} conformation on the duplex or dangling end, respectively, and the $\Delta G$ obtained methylating in \emph{syn} conformation on the relative single strand. Transitions between \emph{syn} and \emph{anti} states were never detected during the alchemical simulations. In this way, the contribution to the free energy given by the \emph{syn} (\emph{anti}) conformation in the duplex (single strand or dangling end) were ignored. Indeed, we expect these contributions to be negligible based on the experimental evidences \cite{roost2015structure,liu2020quantitative}, which show a \emph{syn/anti} isomer preference when paired ($\approx$1:100) versus unpaired ($\approx$10:1).
This was additionally verified with supplementary simulations performed on the A2 and A3 systems.
Moreover, we computed $\Delta G_{syn/anti}$ by performing the alchemical transformations on the isolated nucleoside in solution for the two isomers and computing their difference.

\subsection{Fitting Procedure}
We employ a fitting strategy based on reweighting \cite{cesari2019fitting} where a subset of the partial charges and a dihedral potential are adjusted to match experimental data. Specifically, we decided to fit the atoms that are closer to the methyl group (N6, C6, H61, C10, H101/2/3, and N1,
see Fig.~\ref{fig1}).
The total charge was maintained, leading to 5 free parameters associated to the partial charges.
A single cosine was added to the $\eta_6$
torsional angle identified by atoms N1--C6--N6--C10:
$U(x)  = V_{\eta} [1 + cos(\eta_6 (x) - \pi)]$.
This angle controls the \emph{syn/anti} relative populations, leading to a total of 6 parameters, and the shift is chosen so that a positive
value of $V_{\eta}$ favors \emph{syn} configurations over \emph{anti}.

To optimize the calculation of the total energy of the system at every iteration of our fitting procedure, where up to 6 charges were possibly modified, we notice that the total energy of the system is a quadratic function of the charge perturbations $\Delta Q_i$.
Without loss of generality, one can write the energy change associated to charges and torsion perturbation as
\begin{multline}
\Delta U(x)=\sum_{i=5}^5 K_i(x) \Delta Q_i + \\ \sum_{i=1}^5\sum_{j=i}^5 K_{ij}(x) \Delta Q_i \Delta Q_j + V_{\eta} [1 + cos(\eta_6 (x_i) - \pi)]
\end{multline}
In total, for every analyzed snapshot ($x$), 20 coefficients ($K_i$ and $K_{ij}$) can be precomputed that allows obtaining the energy change
for arbitrary choices of $\Delta Q$ with simple linear algebra operations, without the need to recompute electrostatic interactions explicitly.
The coefficients were obtained by using GROMACS in rerun mode for 20 sets of test charge perturbation
that were extracted from a Gaussian with zero average and variance set to 1 \emph{e}.
The perturbations were constructed to maintain the total charge constant.
Importantly, this approach correctly takes into account the effect of charge perturbations on 1--4 interactions,
where electrostatics is scaled with a force-field-dependent fudge factor, as well as on 1--2 and 1--3 interactions, for which it is discarded,
and interaction with all the periodic images.
The second order expansion above
is exact if one neglects roundoff errors.
The magnitude of charge perturbations was chosen to minimize such errors.
Eq.~\ref{eq:deltag} should then be suitably modified replacing
$\Delta E $ with $\Delta E + \Delta U$.
Its derivatives with respect to the free parameters (charge and dihedral potential coefficient) can be computed as well (see Section S6).

Our fitting is based on the minimization of a L2-regularized cost function defined as follows:
\begin{equation}
\label{cost}
C = \chi ^2 + \alpha \sum_{i=0}^5 \Delta Q_i^2 + \beta V_{\eta}^2
\end{equation}

\begin{equation}
\chi ^2 = \frac{1}{N_{exp}} \sum_{i=0}^{N_{exp}} \frac{(\Delta \Delta G_i^{AFEC} -\Delta \Delta G_i^{exp})^2} {\sigma _i^2}
\end{equation}
Here the $\chi ^2$ measures the discrepancy between computations and experiments, whereas the regularization terms
on the charges and on the torsional $\eta _6$ are governed by the hyperparameters $\alpha$ and $\beta$ and are needed in order to avoid overfitting on the training set.
This function is minimized using the  L-BFGS-B method \cite{zhu1997algorithm} as implemented in SciPy \cite{virtanen2020scipy}.

The result crucially depends on the choice of the hyperparameters  $\alpha$ and $\beta$.
Lower values for the hyperparameters imply that larger corrections are allowed, with the risk of overfitting, and thus lower transferability 
to new experiments.
Higher values for the hyperparameters imply that lower corrections are allowed, with the risk of underfitting, and thus lower accuracy
in reproducing experimental data.
The sweet point could be in principle found with a cross validation (CV) procedure and a scan over possible values for $\alpha$ and $\beta$ \cite{cesari2019fitting,frohlking2020toward}.
For the smallest dataset (set A in Table \ref{table1}), we used a leave-one-out CV strategy, i.e., we trained
the parameters on all systems except one.
For the largest dataset (set AB in Table \ref{table1}), we used a leave-3-out strategy, iteratively training the parameters on 7 randomly chosen experiments and validating on the 3 left out experiments.
In both cases, we then assessed the transferability of the model by evaluating its average  $\chi ^2$ on the system (or the subset of systems) that was left out.

\subsection{Statistical Significance}

When recomputing energies through a reweighting procedure, particular attention must be taken towards the statistical significance that may be  lost during the computation, by reducing the effective sample size of the data set.
This is usually monitored computing the Kish effective sample size \cite{GrayKish1969,rangan2018determination}.
In our case, the most affected ensemble is the one corresponding to m$^6$A ($\lambda=1$).
We thus monitor the Kish size computed using weights corresponding to the $\lambda=1$ ensemble, defined as
\begin{equation}
KS_{\lambda=1} = \frac{[\sum_x w(x) e^{-\beta(\Delta E(x)+\Delta  U(x))}]^2}{\sum_x [w(x) e^{-\beta (\Delta E(x)+\Delta  U(x))}]^2}
\end{equation}
We then compare it with the Kish size obtained with the original force field, defined as
\begin{equation}
KS^0_{\lambda=1} = \frac{[\sum_x w(x) e^{-\beta \Delta E(x)}]^2}{\sum_x [w(x) e^{-\beta \Delta E(x)}]^2}
\end{equation}
To quantify how much statistical efficiency is lost due to the reweighting to a modified set of parameters we
use the Kish size ratio (KSR), that we define as
\begin{equation}
KSR=\frac{KS_{\lambda=1}}{KS^0_{\lambda=1}}
\end{equation}

\section{Results}

In this work, we fit point charges and a single torsional potential correction
for a m$^6$A RNA residue using alchemical MD simulations and a set of experimental
data.
In all the fittings, charges and torsional potential were subject to L2 regularization
with hyperparameters $\alpha$ and $\beta$, respectively.
We initially employed only the first 5 experimental datapoints of Table~\ref{table1},
namely (A1) $\Delta G_{syn/anti}$ for a nucleobase and
(A2--A4) $\Delta \Delta G$ in melting experiments \cite{roost2015structure}.
Thus, we first report the results obtained with such set of charges, including a validation done on
a more recent set of melting experiments (B1--B5) \cite{kierzek2022secondary}.
We then report results obtained with charges that were fitted on the entire dataset (A1--A5 and B1--B5).
As a reference, results obtained with the Aduri \emph{et al.} \cite{aduri2007amber} modifications (modrna08) for the commonly
used AMBER force field are also reported,
either \emph{as is} or complemented with a custom torsional correction that result in a $\Delta G_{syn/anti}$ matching  experiment A1.
All the calculated $\Delta G$s are reported in Table S3.
A complete list of the performed alchemical simulations is reported in Section S9.

\subsection{Fitting on the smaller dataset}

For this first fitting, we only employed data set A1--A5 from Table~\ref{table1}.
$\chi ^2$ errors were computed assuming the experimental error of each data point to be equal to each other and to
1 kJ/mol.

\begin{figure}
\begin{center}
\includegraphics[width=1.0\textwidth]{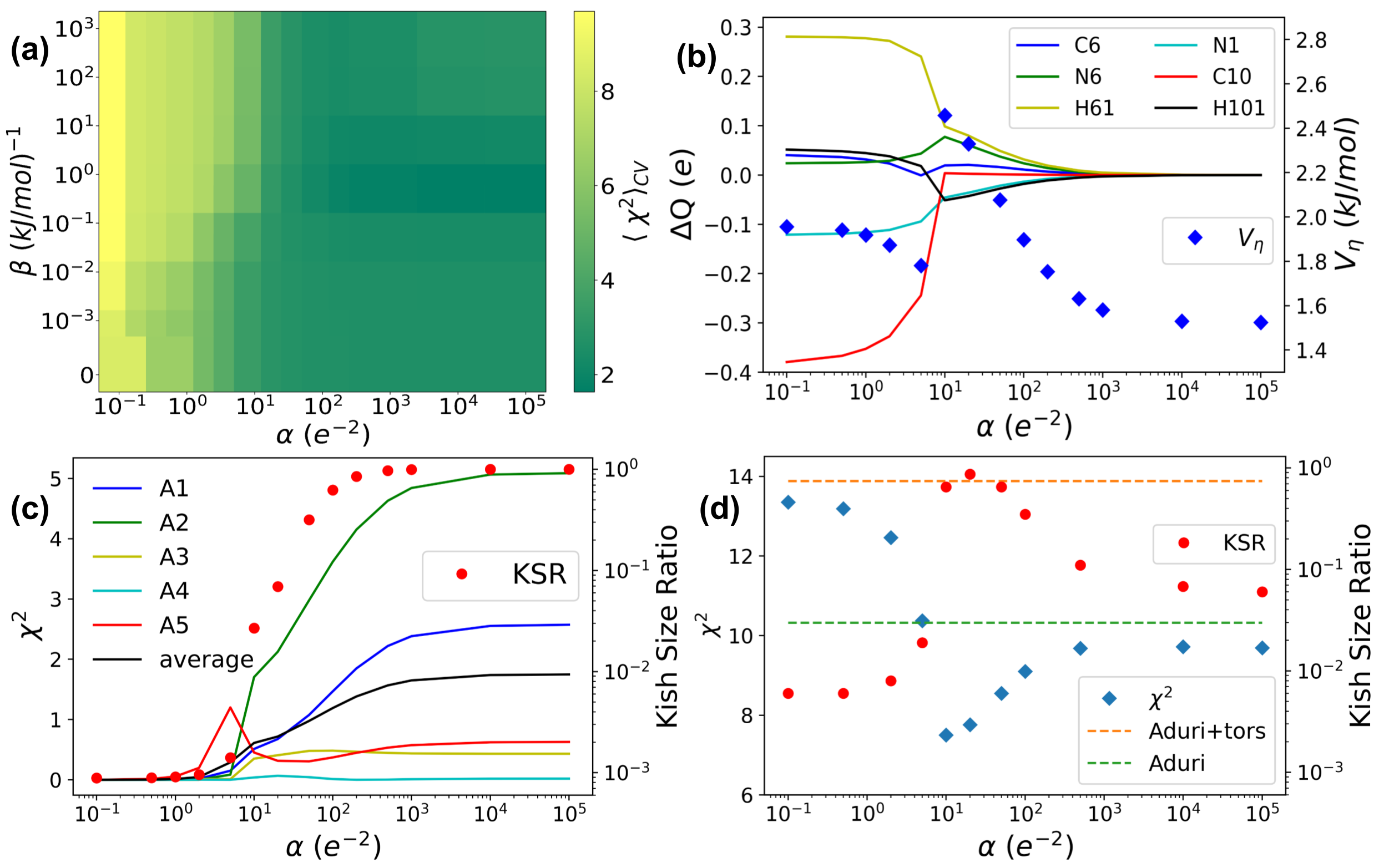}
\end{center}
\caption{Results obtained with parameters fitted on the initial dataset, A1--A5 in Table \ref{table1}.
Cross validation error obtained with a leave-one-out-procedure, shown as a function of the
two regularization hyperparameters $\alpha$, for charges, and $\beta$, for the torsional potential (panel a).
Darker green colors correspond to lower values of the average $\chi ^2$ computed on the systems left out iteratively from the fitting.
Parameters ($\Delta Q$ and $V_{\eta}$) obtained from the entire initial dataset
as a function of $\alpha$, with $\beta=0$ (panel b).
$\chi^2$ errors for individual experiments
and Kish size ratio (KSR, see text for definition) obtained
using parameters fitted on the entire initial dataset 
as a function of $\alpha$, with $\beta=0$ (panel c).
Validation on the second dataset (B1--B5 in Table \ref{table1}) of the parameters obtained on the first dataset  (panel d).
Results using Aduri parameters are shown as horizontal lines, either as reported in the original paper (green) or including a single torsional correction
to obtain the correct \emph{syn/anti} population (data point A1).}
\label{figure_fit1}
\end{figure}

Figure~\ref{figure_fit1}a reports the results of a cross-validation test performed with a leave-one-out procedure.
Namely, we fit the whole experimental dataset leaving out one experimental data point at a time and report
the average error on the left-out experiment. In this leave-one-out procedure
we decided not to iterate on the $\Delta G_{syn/anti}$ experiment (A1), since this is expected to be crucial to correctly
reproduce the conformation of non-Watson--Crick-paired residues (mostly \emph{syn}).
From this map, we can hardly appreciate any variation of the $\chi ^2$ along the vertical axis corresponding to the $\beta$ hyperparameter.
This suggests that $\beta$ could be set to zero, thus simplifying all subsequent hyperparameter scans.
Conversely, the $\chi ^2$ grows significantly for low $\alpha$ values.
This implies that regularization on charges is required to avoid overfitting.
In general, one should expect a minimum to be observed in this type of hyperparameter scans \cite{cesari2019fitting,frohlking2020toward}.
This is not the case here for the $\alpha$ scan (see also Fig.~S3,
showing projection on $\alpha$ for $\beta=0$), implying that the performance of the parameters on a given system are not improved when excluding
that system from the training set. This is likely due to the small data set employed.

Figure~\ref{figure_fit1}b shows the optimized parameters (charge and torsional corrections) as a function of the regularization hyperparameter $\alpha$
while fixing $\beta=0$.
A transition can be seen at $\alpha \approx 10$. Namely, 
when $\alpha>10$, parameters have a smooth dependence on $\alpha$, whereas 
when $\alpha<10$, both the charges and the torsional potential change suddenly.
In the limit $\alpha \rightarrow \infty$, it can be seen that charge corrections tend to zero with an inverse law dependence, which is expected for L2 regularization, 
and the torsional correction tends to $V_{\eta}\approx$ 1.5 kJ/mol, which corresponds
to the amplitude of the torsional potential that optimizes the $\chi ^2$ without modifying the charges of the reference Aduri \emph{et al.} model.
We notice that $\Delta G_{syn/anti}$ obtained when using the Aduri \emph{et al.} force field is $\approx$ 1.7 kJ/mol, and thus this correction
results in $\Delta G_{syn/anti}\approx 1.7 + 2 \times 1.5=4.7 $  kJ/mol, which is still smaller than the experimental reference $\approx$ 6 kJ/mol.
The obtained parameters indeed strike a balance between favoring the \emph{syn} state in the isolated nucleoside and
not favoring it too much in the single-stranded calculations used to predict the $\Delta \Delta G$ from melting experiments,
which would lead to too large destabilizations associated to the methylation.
When $\alpha$ is decreased, the optimal torsional correction changes,
since all the parameters are coupled. This confirms that charges and torsional parameters should be
fitted simultaneously.

Figure~\ref{figure_fit1}c shows the individual $\chi ^2$ associated to the same hyperparameter scan.
The average $\chi^2$ error is, by construction, monotonically increasing with $\alpha$,
and most of the individual errors follow the same trend.
Figure~\ref{figure_fit1}c also shows the statistical efficiency of the analysis, quantified by the relative reduction
of the Kish effective sample size associated to reweighting.
A low number here indicates that the tested charges are so different from those employed in the simulation
to make the result statistically not significant.
The Kish size displays a significant drop for $\alpha<10$, indicating that results in this regime
might be not significant. This is a likely explanation for the discontinuous behavior observed in Fig.~\ref{figure_fit1}b.

We then tested the charges obtained with this reduced training set on the newer data set B1--B5, see Table \ref{table1}, which was not included
in the training phase.
This set of data involves 5 recently published melting experiments \cite{kierzek2022secondary},
4 of which have m$^6$A occurring in both chains of the duplex.
We notice that double methylations are expected to lead to an even lower statistical efficiency of the reweighting procedure.
We thus performed this analysis by reweighting simulations that were generated using an alternative
set of parameters that was derived with a similar fitting procedure shown in this paper, but with an incorrect regularization on the charges as described in Section S4. Since this parametrization is closer to the right solution of the fitting with respect to Aduri, it
obtains higher Kish size values in the relevant $\alpha$ range (see Fig.~\ref{figure_fit1}d).
The $\chi ^2$ computed on the second data set shows that an optimal result can be obtained setting $\alpha\approx 10$.
We also compared with results obtained using the original Aduri charges and optionally including a torsional correction
to fix the \emph{syn/anti} balance.
These results are obtained with direct simulation, that is without reweighting.
It can be seen that the results with the parameters trained on systems A1--A5 largely outperform those obtained with Aduri
parameters on systems B1--B5, thus confirming the transferability of the parameters.
Aduri+tors parametrization corresponds to setting $V_{\eta}$ =2.35 kJ/mol in such a way to perfectly fit experiment A1 (single nucleoside) without modifying charges. $\chi^2$ computed for Aduri+tors demostrates that acting exclusively on the torsional is not sufficient to reproduce both $\Delta G_{syn/anti}$ and melting experiments.
It is also important to note that the improvement in reproducing experiments is obtained by changes in the partial charges
that are small when compared to differences between charges derived
with the standard restrained electrostatic potential protocol \cite{cieplak1995application} in different conformations (see Section S5).

\subsection{Fitting on the full data set}

Next, we perform a fitting using the full data set reported in Table~\ref{table1}.
Since the variability of error in this data set is larger, we here computed $\chi^2$ using the experimental errors reported in Table~\ref{table1}.
For the  $\Delta G_{syn/anti}$ experiment, for which an experimental error is not reported, we used a nominal $\sigma=0.5$ kJ/mol
so as to assign to this experiment a larger weight when compared to the other data points corresponding to melting experiments.

\begin{figure}
\begin{center}
\includegraphics[width=1.0\textwidth]{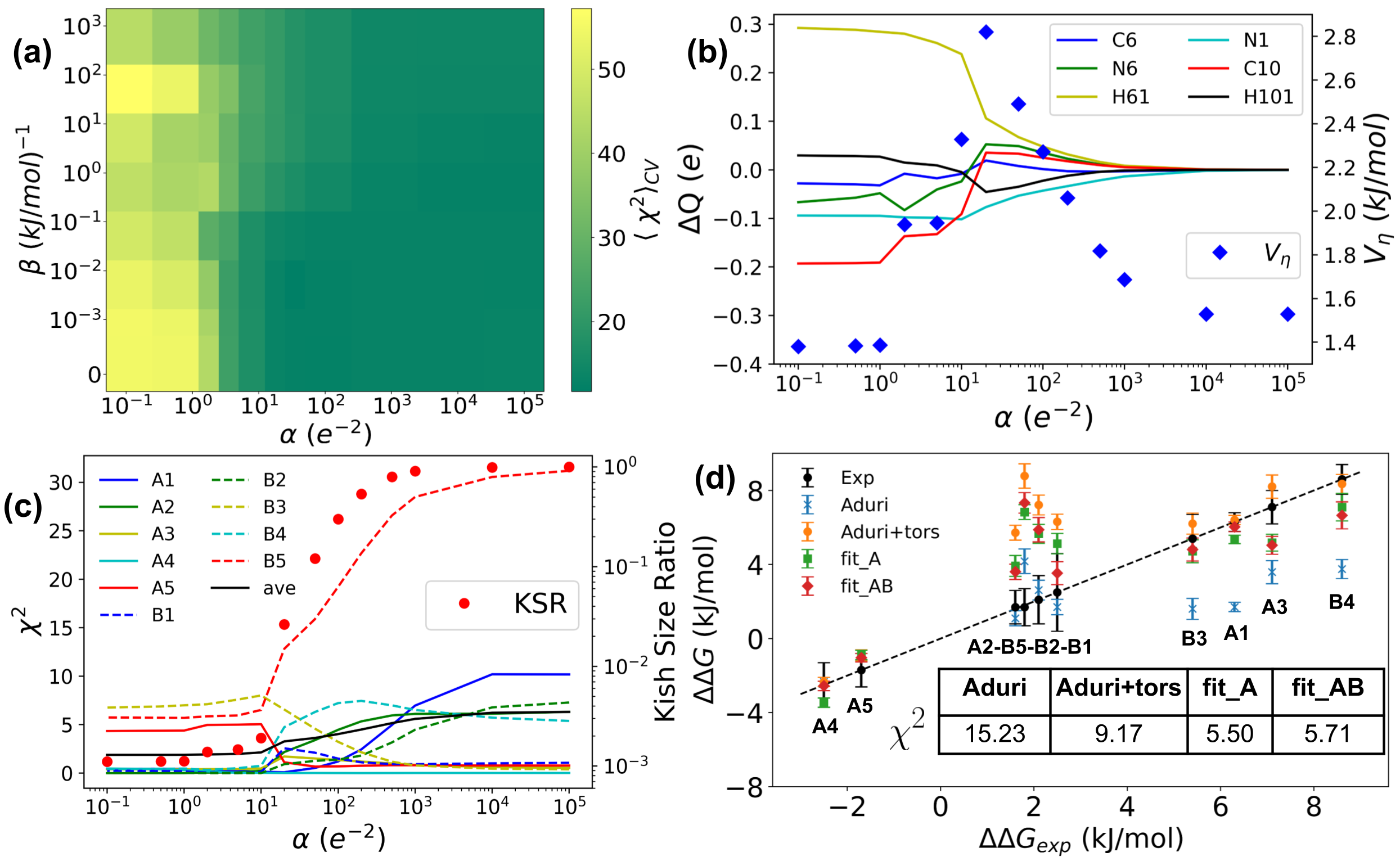}
\end{center}
\caption{Results obtained with parameters fitted on the full dataset, A1--A5 and B1--B5 in Table \ref{table1}.
Cross validation error obtained with a leave-three-out-procedure, shown as a function of the
two regularization hyperparameters $\alpha$, for charges, and $\beta$, for the torsional potential (panel a).
Darker green colors correspond to lower values of the average $\chi ^2$ computed on the systems left out iteratively from the fitting.
Parameters ($\Delta Q$ and $V_{\eta}$) obtained from the entire dataset
as a function of $\alpha$, with $\beta=0$ (panel b).
$\chi^2$ errors for individual experiments
and Kish size ratio (KSR, see text for definition)
using parameters fitted on the entire initial dataset 
as a function of $\alpha$, with $\beta=0$ (panel c).
$\Delta \Delta G$ computed for each of the ten analyzed systems with 4 different sets of parameters
(panel d).
fit\_A are parameters obtained fitting on the first data set (A1--A5) with regularization $\alpha=10$.
fit\_AB are derived fitting on the entire data set (A1--A5 and B1--B5) for $\alpha=50$.
$\chi ^2$ obtained for each force field set of parameters are shown in the table inside panel d.}
\label{figure_fitAB}
\end{figure}

Figure~\ref{figure_fitAB}a reports the results of a cross-validation test performed with a leave-three-out procedure.
Namely, we randomly select seven systems to be used in training and we report the average $\chi ^2$ error
obtained for the remaining three systems. This time also system A1 was allowed to be left out from the training set.
Results are qualitatively consistent with those obtained with the smaller data set (see Fig.~\ref{figure_fit1}).
It is difficult to appreciate any variation of the $\chi ^2$ along the vertical axis corresponding to the $\beta$ hyperparameter, suggesting that we can safely set $\beta=0$.
We also do not find any clear minimum when scanning over $\alpha$
(see also Fig.~S4, showing projection on $\alpha$ for $\beta=0$).
Figure~\ref{figure_fitAB}b shows the parameters as a function of the regularization hyperparameter $\alpha$ 
while fixing $\beta$.  A clear transition can be seen at $\alpha \approx 20$. 
The average $\chi^2$ error is monotonically increasing with $\alpha$,
but some of the systems have a non-trivial behavior (Fig.~\ref{figure_fitAB}c).
The Kish size shows a significant drop for $\alpha<50$, showing that results in this regime
might be not statistically reliable.
We thus select the parameters obtained with $\alpha=50$ as the optimal ones trained on the entire data set.

We then compare the performance of a number of different sets of parameters in reproducing all the available experimental data points.
Namely we compare (a) the original Aduri parameters (Aduri), (b) the Aduri parameters augmented with a torsional correction to enforce the
correct \emph{syn/anti} balance in a nucleobase (Aduri+tors), (c) the parameters obtained fitting on the initial dataset (A1--A5), with hyperparameter $\alpha=10$ (fit\_A), 
and (d) the parameters obtained fitting on the full dataset (A1--A5 and B1--B5), with hyperparameter $\alpha=50$ (fit\_AB).
Results are reported in Fig.~\ref{figure_fitAB}d. The quality of the fit is also summarized in the reported $\chi^2$ values.
The addition of a simple torsional correction to the Aduri parameters results in an overall
worsening of the agreement with experiment. On the other hand, the two sets of parameters obtained in this work (fit\_A and fit\_AB)
display a significantly better agreement with experimental data. Points are computed directly from energies generated during the MD, except for fit\_A case in which energies are computed through a reweighting procedure that gives a KSR about 0.65.
Note that even if fit\_A has the best performance, fit\_AB allows to get very similar $\chi ^2$ with a stronger regularization (higher $\alpha$), keeping the parametrization closer to the reference one, as shown in Figs.~S6--S7.
Finally, as observed in the previous subsection, we note that the improvement in reproducing experiments is obtained by relatively small changes in the partial charges (see also Section S5). The fitted parameters are summarized in Table \ref{tablePar}.

\begin{table}
\begin{center}
\resizebox{\columnwidth}{!}{%
\begin{tabular}{|l|c|c|c|c|c|c|r|}
\hline
      & \textbf{C6} (e) & \textbf{N6} (e) & \textbf{H61} (e) & \textbf{N1} (e) &\textbf{C10} (e) & \textbf{H100} (e) & \textbf{$V_{\eta}$} (kJ/mol)\\
\hline
\textbf{fit\_A} & 0.019 & 0.077 & 0.099 & -0.046 & 0.004 & -0.051 & 2.46 \\
\hline
\textbf{fit\_AB} & 0.009 & 0.049 & 0.067 & -0.053 & 0.033 & -0.035 & 2.49\\
\hline

\end{tabular}%
}
\caption{Charge modifications ($\Delta Q$s) and torsional potential ($V_{\eta}$) for the fitting performed on the smaller dataset (fit\_A, $\alpha = 10$)
and for the fitting performed on the larger dataset (fit\_AB, $\alpha = 50$).} 
\label{tablePar}
\end{center}
\end{table}

\subsection{Relative stability of \emph{syn} and \emph{anti} conformations}
One piece of the experimental information that we implicitly used in our fitting procedure is the relative stability of \emph{syn} and \emph{anti} conformations
in a nucleoside (data point A1, in Table \ref{table1}).
We indeed assumed a predominant population of \emph{syn} conformation for the unpaired nucleotides used in the reference single-stranded systems.
We also assumed that m$^6$A adopts exclusively its \emph{anti} conformation when paired, in agreement with experiments \cite{roost2015structure,liu2020quantitative}.
In particular, Ref.~\cite{liu2020quantitative} reports that, for the most common G6C sequence,
m$^6$A forms a Watson--Crick base pair with uridine that transiently exchanges on the millisecond time-scale
between a main substate (\emph{anti}) and a low populated (1\%), singly hydrogen-bonded and mismatch-like conformation through isomerization of
the methylamino group to the \emph{syn} conformation.
This population corresponds to a
$\Delta G_{syn/anti}^{duplex} \approx -11$ kJ/mol.
We \emph{a posteriori} validated this population by performing alchemical transformations on the duplex systems enforcing the
\emph{syn} conformation. The predicted $\Delta G_{syn/anti}$ for a nucleotide and for two of the tested duplexes are reported
in Table \ref{table2}, where the corresponding experimental values are also included.
For the A1 experiment, as expected, the proposed sets of parameters closely match the experimental value, that was
used during training. The Aduri \emph{et al.} force field underestimates the $\Delta G_{syn/anti}$, resulting
in a relatively high population of the unexpected \emph{anti} conformation in a nucleoside.
This difference can be directly corrected with a torsional potential applied on the $\eta$ torsion
(Aduri+tors).
However, when analyzing duplexes A2 and A3 with the Aduri+tors parameters,
we found that the predicted  $\Delta G _{syn/anti}$ would be close to zero, in fact
making the assumption of neglecting the \emph{syn} conformation in duplexes in our
alchemical calculations difficult to justify, and in disagreement with experimental findings.
In other words, the original Aduri charges
allow to reproduce the relative stability of \emph{syn} and \emph{anti} conformations either in the paired state (Aduri parameters)
or the unpaired state (with torsional correction), but not in both simultaneously.
Remarkably, the sets of parameters proposed here, that also contain a torsional
term penalizing the \emph{anti} conformation, result in a significantly higher value for $\Delta G _{syn/anti}^{dup}$,
much closer to a qualitative agreement with experiment.
This suggests that the proposed parameters better describe the interactions of the m$^6$A nucleobase with the surrounding environment
and are thus more transferable.
We notice that the relative stability of \emph{syn} and \emph{anti} conformations is predicted to be sequence dependent,
being different for system A3 (sequence U6G).

\begin{table}
\resizebox{\columnwidth}{!}{%
\begin{tabular}{|l|c|c|c|c|r|}
\hline
\multicolumn{1}{|c|}{\textbf{}}            &            \textbf{Aduri}            & \textbf{Aduri+tors}   & \textbf{fit\_A}     &    \textbf{fit\_AB}  & \multicolumn{1}{c|}{\textbf{Exp}}\\
\hline
A1. $\Delta G_{syn/anti}$ & 1.71 $\pm$ 0.25  & 6.33 $\pm$ 0.25 & 5.66  $\pm$ 0.21 & 6.04 $\pm$ 0.26  & 6.3  \\
\hline
A2. $\Delta G_{syn/anti}^{dup}$           &           - 7.7   $\pm$ 0.5       &            - 3.1 $\pm$ 0.4       &  - 9.7  $\pm$ 0.5  &            - 7.8   $\pm$ 0.4  & $\sim$ -11 \\
\hline
A3. $\Delta G_{syn/anti}^{dup}$           &            - 5.4 $\pm$ 0.5          &            - 0.8 $\pm$ 0.4     &  - 5.1  $\pm$ 0.4     &            - 5.8 $\pm$ 0.5  & -- \\
\hline
\end{tabular}%
}
\caption{Free-energy differences between \emph{syn} and \emph{anti} isomer states in systems A1--A3. The last column corresponds to experimental estimates, whereas other columns corresponds to computed $\Delta \Delta G$ for different parametrization. Energies are given in kJ/mol units. }
\label{table2}

\end{table}

\begin{figure}
\includegraphics[width=1.\linewidth]{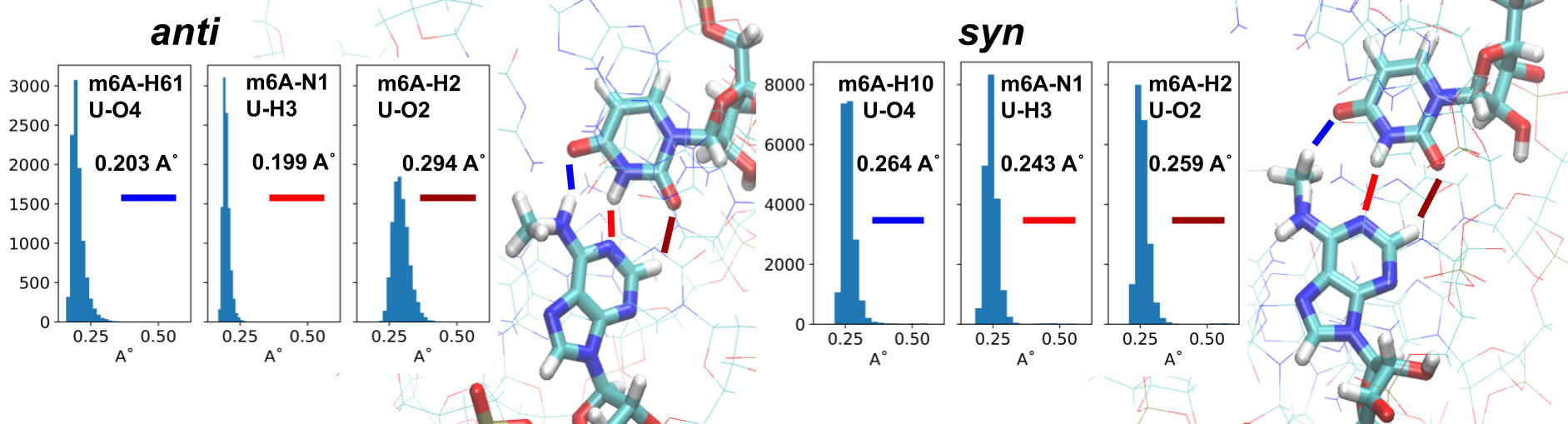}
\caption{Interfacing atom distances for m$^6$A-U pairing in system A2 in Table \ref{table1}, for \emph{anti} conformation (left) and \emph{syn} (right). Histograms show unimodal distributions, and the averaged value are indicated in the box. Distances are sampled from the alchemical trajectories considering only the $\lambda = 1$ replica. In the \emph{syn} conformation, m$^6$A-H10 correspond to the hydrogen of the methyl group closest to the uracil oxygen O4.}
\label{fig5}
\end{figure}

To gain insight about how the m$^6$A--U pairings occurs in the duplexes, we analyzed snapshots of system A2, both for m$^6$A in \emph{syn} and \emph{anti}, together with histograms of distances between atoms belonging to the two nucleobases (Fig.~\ref{fig5}).
The reported histograms are unimodal and with an increased average associated to the distortion of the  A-U Watson--Crick pairings due to the steric clash induced by the methylation.
However, the hydrogen bond between A-N1 and U-H3 is present, in agreement with what has been suggested previously~\cite{liu2020quantitative}.

\subsection{Interpretation of parameters and dependence on ionic strength}

To provide an interpretation for the obtained parameters, we performed a few additional fittings. In particular we investigated which charges have a major impact on enforcing agreement with experiments. To this purpose we performed fitting on exclusively two charges at a time plus the torsional term, considering couples of charges that in fit\_A and fit\_AB have systematically positive and negative $\Delta Qs$. Results are discussed in Section S7 and Fig.~S9.
Overall, the results suggest that the main contribution of the fitted correction is to increase the stability of Watson--Crick hydrogen bonds
by making N1 and H61 more polar and at the same time using the $\eta$ torsional potential to control the \emph{syn}/\emph{anti} relative population.

As a further test, we simulated a subset of the systems using a salt concentration of 1 M, which is consistent with that used in experiments. As shown in Section S8, results are equivalent to those obtained at 0.1 M salt concentration. 
We interpret this result with the fact that the methylation is not altering sufficiently the electrostatic environment to be sensitive to changes in ion concentration.
This implies that a training performed using simulations performed at a different ion concentration would result in an equivalent set of parameters and further confirms the robustness of our results.

\section{Discussion}

In this work, we proposed a protocol to parametrize charges in modified nucleobases using available melting experiments.
The approach is applied to m$^6$A and leads to a set of charges that can reproduce a set of 10 independent experimental values.
The approach is based on the force-field fitting strategies introduced in earlier works \cite{norgaard2008experimental,li2011iterative,cesari2019fitting}, which
are here extended with a number of technical improvements.

A first methodological contribution is a formalism that allows alchemical calculations to be used as a reference.
Previous works
were only using observables computed with a single set of force-field parameters
\cite{norgaard2008experimental,li2011iterative,cesari2019fitting,frohlking2022automatic}.
The method introduced here allows free-energy differences between different sets of parameters to be evaluated
and compared with experiment. This opens the way to the optimization of parameters based on experimentally measured $\Delta \Delta G$s.
We based our analysis on optical melting experiments, which are commonly employed in the nucleic acids community \cite{schroeder2009optical},
but other types of experiments might be considered.
In our specific application, only the parameters of one of the two end states were refined, but one could similarly fit
parameters for both adenosine and m$^6$A, at the price of increasing the number of parameters and thus the risk of overfitting.
A second improvement is that we developed a way to efficiently recompute the total energy of the system
using test charges. This is achieved by precomputing the total electrostatic energy of the system with a set
of randomly perturbed charges. Given the high cost of electrostatic calculations, this makes the cost of each of the iterations
performed during force-field fitting significantly faster, and implicitly takes into account combination rules, non-bonded exclusions, and periodicity.
These two improvements can be readily integrated in other MD-based force-field optimization strategies.

A limitation of optimizing charges with the introduced procedure is that the statistical efficiency of reweighting is significantly decreased even by
small charge perturbations.
This implies that simultaneously parametrizing many copies of the same nucleotide,
or parametrizing a larger number of charges for the same nucleotide, would be more difficult.
In our case, we had to include at most two m$^6$A residues in the same simulation.
If more copies of the same reparametrized nucleotide are needed in the same system,
one might have to design strategies where only a few copies at a time are reparametrized, or follow an iterative procedure where
modifications are included in consecutive steps \cite{cesari2019fitting}. In this application, this was not necessary.

Overfitting was avoided by using a standard L2 regularization term on the charge increments.
This penalty does not depend the charge location.
Importantly, the regularization hyperparameters tune the relative weight of the experimental data and of
the reference charges, here taken from Ref.~\cite{aduri2007amber}, thus allowing to achieve
a meaningful set of parameters also in regimes where the number of data points is very limited.
More effective regularization strategies might
be designed based on the molecular dipole, as done in Ref.~\cite{janecek2021w}, so as to minimize the perturbation of the electrostatic potential
at a large distance from the molecule. Alternatively, one might directly use as a regularization term the deviation from the quantum-mechanical electrostatic potential
at short distance.
In the limit of a large regularization hyperparameter, 
this would lead to ESP charges \cite{singh1984approach}.
Finally, other regularization criteria might be used \cite{frohlking2022automatic}.
When comparing our procedure with standard ESP charge fitting, it is important to realize that we are aiming to reproduce experimentally
observed $\Delta \Delta G$, which are non-linear functions of the energy of each configuration, which in turn is a quadratic function
of the charges. These non-linearities make it possible for multiple local minima of the cost function to exist, and could thus make 
the minimization not reproducible. However, when sufficiently regularized, the fitting procedure results in reproducible charges that depend smoothly
on the control parameters.
In standard ESP fitting, instead, the electrostatic potential is fitted, thus resulting in a linear fit with a unique solution.

We notice that the parameters of the unmethylated force field were not modified. This was based on the assumption that
the employed set of force-field parameters is already capable to reproduce $\Delta \Delta G$ experiments associated to mutations
between non-modified nucleobases \cite{Kameda2015WCenergies}. The m$^6$A charge optimization could be easily repeated using another set of initial parameters,
and the parameters of non-modified nucleobases might be adjusted as well, although with the caveat discussed above.

Another possible limitation of the employed alchemical simulations is the sufficient sampling of the end states.
The duplex is expected to be stable and well structured, so that sampling multiple structures should not be necessary.
For selected cases, we also explored the possibility to include the unlikely \emph{syn} paired state, which, as expected,
gives a negligible contribution to the stability of the duplex. For single strands, instead, we only sampled the \emph{syn} state.
More importantly, our simulations were short enough to avoid any significant reconformation of the single strand.
Sampling the conformations of flexible, single stranded RNAs is notoriously difficult \cite{sponer2018rna}.
In addition, the generated ensemble might contain artificially stabilized intercalated structures, whose population
is known to be overestimated by the RNA and water force fields adopted here \cite{condon2015stacking,bottaro2018conformational}.
This would make the correct sampling of the single stranded state unfeasible.
We also notice that the experimental results that we aimed to reproduce were performed on systems designed to have
the isolated strands unstructured, so as to capture the effect of methylation on hybridization.
Putting everything together, we conclude that the approximation of a single strand ensemble that do not depart too much
from the initial A-form helix is a sensible choice for this specific application.

An important finding of this work is that the parameters of Aduri \emph{et al.} cannot reproduce the \emph{syn}/\emph{anti}
balance expected for m$^6$A residues. This balance is extremely important, and is related to the mechanism by which m$^6$A modifications
modulate duplex stability \cite{roost2015structure}. This could not be rectified with a straightforward correction on the single involved torsion.
The optimized charges, instead, allow the correct \emph{syn}/\emph{anti} balance to be recovered both in paired and unpaired nucleobases,
as well as an heterogenous set of optical melting experiments to be reproduced.
Interestingly, the Aduri \emph{et al.} parameters were tested in a recent work \cite{hurst2021deciphering}, with
results for system A2 in Table \ref{table1} consistent with ours and with experiments.
However, systems A1 and A3 were not tested, and thus the problem that we observed here could not be identified.
Another interesting finding is that the  $\Delta \Delta G$ associated to N6 methylation are here predicted to be independent of ion concentration. We are not aware of any experimental validation of this finding, which could be obtained by comparing melting experiments at different ion concentration.
Finally, our results suggest that the relative population of the \emph{syn} excited state in duplexes \cite{liu2020quantitative}
might significantly depend on the identity of the neighboring nucleotides. The precise hybridization kinetics could thus be quantitatively different
for RNAs with different sequences.

A convenient property of our approach is that it does not require changing the functional form of the
interaction potential, so that new parameters can be readily incorporated in existing MD software. This is not the case if \emph{ad hoc} corrections
are employed \cite{kuhrova2019improving,frohlking2022automatic}. In addition, it is worth noting that the charge modifications
obtained are very small, and in particular they are
smaller than the typical difference between sets of charges derived with slightly different procedures or using different
reference conformations. In spite of this small difference, the effect on experimental observables is significant.
These observations imply that there is still significant space to improve the performance of current force fields without
necessarily modifying the functional form, if experimental information is used during training \cite{frohlking2020toward}.

Using our approach it is possible to dissect the individual contribution of the modified force-field parameters.
The main factors playing a role on the change of duplex stability induced by m$^6$A methylation are
(a) the penalty for switching to the unfavored \emph{anti} isomer \cite{roost2015structure},
(b) the stabilization induced by hydrophobic shielding of the methyl group against surrounding bases (see also Fig.~S8) \cite{guckian2000factors,isaksson2005uniform},
(c) the impact of partial charges on stacking interactions \cite{guckian2000factors}, and
(d) the impact on the strength of Watson--Crick hydrogen bonds. 
Since, on average, experimental $\Delta \Delta G$ for denaturation experiments performed on duplexes are smaller than the \emph{anti} isomer penalty, we could expect that the sum of the other factors has a stabilizing effect on the majority of the considered duplexes.
We notice that Aduri charges for N1 and H61, which are involved in Watson--Crick pairings with the complementary uridine, have partial charge absolute values significatively
lower compared to the standard adenine parameters (0.28948 vs 0.41150 for H61, $-0.675968$ vs. $-0.76150$ for N1).
This may lead to a weakening of hydrogen bonds which may cause an overestimation of destabilization induced on duplexes, as we observed in Aduri+tors cases (see Fig.~\ref{figure_fitAB}d). The results of our fitting systematically increase the absolute value of H61 and N1 partial charges, hence resulting in a stronger Watson--Crick pairing. At the same time, the torsional term allows to reproduce the correct \emph{anti} isomer penalty. Parameters are coupled, so that it is necessary to fit them simultaneously so as to avoid double counting effects.

To the best of our knowledge, this is the first attempt to tune partial charges of a biomolecular force field based on experiments performed on
macro-molecular complexes. We expect that this approach could be used in the future to improve the capability of biomolecular force fields
to match experimental observations exploiting a part of the functional form that has been traditionally derived in a bottom-up fashion.
Remarkably, the parameters derived here for m$^6$A allow to 
properly describe paired and unpaired m$^6$A in both \emph{syn} and \emph{anti} conformation,
and thus open the way to the use of molecular simulations to quantitatively investigate the effects of N6 methylations on RNA structural dynamics.

\section{Data availability}

Jupyter notebooks used for molecular dynamics simulations and analysis can be found at \url{https://github.com/bussilab/m6a-charge-fitting}.
Input files and trajectory data are available at \url{https://doi.org/10.5281/zenodo.6498020}.

\section*{Acknowledgments}

David H.~Mathews is acknowledged for reading the manuscript and providing useful suggestions.

\bibliographystyle{unsrt}
\bibliography{main}%

\begin{thebibliography}{10}

\bibitem{gilbert2016messenger}
W.~V. Gilbert, T.~A. Bell, and C.~Schaening.
\newblock Messenger {RNA} modifications: form, distribution, and function.
\newblock {\em Science}, 352(6292), 1408--1412, (2016).

\bibitem{harcourt2017chemical}
E.~M. Harcourt, A.~M. Kietrys, and E.~T. Kool.
\newblock Chemical and structural effects of base modifications in messenger
  {RNA}.
\newblock {\em Nature}, 541(7637), 339--346, (2017).

\bibitem{patil2016m}
D.~P. Patil, C.-K. Chen, B.~F. Pickering, A.~Chow, C.~Jackson, M.~Guttman, and
  S.~R. Jaffrey.
\newblock {m6A} {RNA} methylation promotes {XIST}-mediated transcriptional
  repression.
\newblock {\em Nature}, 537(7620), 369--373, (2016).

\bibitem{he2021m6a}
P.~C. He and C.~He.
\newblock {m6A} {RNA} methylation: from mechanisms to therapeutic potential.
\newblock {\em EMBO J.}, 40(3), e105977, (2021).

\bibitem{tanzer2019rna}
A.~Tanzer, I.~L. Hofacker, and R.~Lorenz.
\newblock {RNA} modifications in structure prediction--status quo and future
  challenges.
\newblock {\em Methods}, 156, 32--39, (2019).

\bibitem{liu2015n}
N.~Liu, Q.~Dai, G.~Zheng, C.~He, M.~Parisien, and T.~Pan.
\newblock N 6-methyladenosine-dependent {RNA} structural switches regulate
  {RNA}--protein interactions.
\newblock {\em Nature}, 518(7540), 560--564, (2015).

\bibitem{huang2017control}
L.~Huang, S.~Ashraf, J.~Wang, and D.~M. Lilley.
\newblock Control of box {C/D} {snoRNP} assembly by {N6}-methylation of
  adenine.
\newblock {\em EMBO Rep.}, 18(9), 1631--1645, (2017).

\bibitem{jones2022structural}
A.~N. Jones, E.~Tikhaia, A.~Mour{\~a}o, and M.~Sattler.
\newblock Structural effects of {m6A} modification of the {X}ist {A}-repeat
  {AUCG} tetraloop and its recognition by {YTHDC1}.
\newblock {\em Nucleic Acids Res.}, 50(4), 2350--2362, (2022).

\bibitem{roost2015structure}
C.~Roost, S.~R. Lynch, P.~J. Batista, K.~Qu, H.~Y. Chang, and E.~T. Kool.
\newblock Structure and thermodynamics of {N6}-methyladenosine in {RNA}: a
  spring-loaded base modification.
\newblock {\em J. Am. Chem. Soc.}, 137(5), 2107--2115, (2015).

\bibitem{hopfinger2020predictions}
M.~C. Hopfinger, C.~C. Kirkpatrick, and B.~M. Znosko.
\newblock Predictions and analyses of {RNA} nearest neighbor parameters for
  modified nucleotides.
\newblock {\em Nucleic Acids Res.}, 48(16), 8901--8913, (2020).

\bibitem{liu2020quantitative}
B.~Liu, H.~Shi, A.~Rangadurai, F.~Nussbaumer, C.-C. Chu, K.~A. Erharter, D.~A.
  Case, C.~Kreutz, and H.~M. Al-Hashimi.
\newblock A quantitative model predicts how {m6A} reshapes the kinetic
  landscape of nucleic acid hybridization and conformational transitions.
\newblock {\em Nat. Commun.}, 12, 5201, (2021).

\bibitem{sponer2018rna}
J.~{\v S}poner, G.~Bussi, M.~Krepl, P.~Ban{\'a}{\v s}, S.~Bottaro, R.~A. Cunha,
  A.~Gil-Ley, G.~Pinamonti, S.~Poblete, P.~Jure\v{c}ka, N.~G. Walter, and
  M.~Otyepka.
\newblock {RNA} structural dynamics as captured by molecular simulations: a
  comprehensive overview.
\newblock {\em Chem. Rev.}, 118(8), 4177--4338, (2018).

\bibitem{tan2018rna}
D.~Tan, S.~Piana, R.~M. Dirks, and D.~E. Shaw.
\newblock {RNA} force field with accuracy comparable to state-of-the-art
  protein force fields.
\newblock {\em Proc. Natl. Acad. Sci. U.S.A.}, 115(7), E1346--E1355, (2018).

\bibitem{frohlking2022automatic}
T.~Fr{\"o}hlking, V.~Ml{\`y}nsk{\`y}, M.~Jane{\v{c}}ek, P.~K{\"u}hrov{\'a},
  M.~Krepl, P.~Ban{\'a}{\v{s}}, J.~{\v{S}}poner, and G.~Bussi.
\newblock Automatic learning of hydrogen-bond fixes in an {AMBER} {RNA} force
  field.
\newblock {\em arXiv preprint arXiv:2201.04078}, (2022).

\bibitem{xu2016additive}
Y.~Xu, K.~Vanommeslaeghe, A.~Aleksandrov, A.~D. MacKerell~Jr, and L.~Nilsson.
\newblock Additive {CHARMM} force field for naturally occurring modified
  ribonucleotides.
\newblock {\em J. Comput. Chem.}, 37(10), 896--912, (2016).

\bibitem{li2019flexible}
Y.~Li, R.~K. Bedi, L.~Wiedmer, D.~Huang, P.~Sledz, and A.~Caflisch.
\newblock Flexible binding of {m6A} reader protein {YTHDC1} to its preferred
  {RNA} motif.
\newblock {\em J. Chem. Theory Comput.}, 15(12), 7004--7014, (2019).

\bibitem{li2021atomistic}
Y.~Li, R.~K. Bedi, L.~Wiedmer, X.~Sun, D.~Huang, and A.~Caflisch.
\newblock Atomistic and thermodynamic analysis of {N6}-methyladenosine (m6a)
  recognition by the reader domain of {YTHDC1}.
\newblock {\em J. Chem. Theory Comput.}, 17(2), 1240--1249, (2021).

\bibitem{hurst2021deciphering}
T.~Hurst and S.-J. Chen.
\newblock Deciphering nucleotide modification-induced structure and stability
  changes.
\newblock {\em RNA Biol.}, 18(11), 1920--1930, (2021).

\bibitem{krepl2021recognition}
M.~Krepl, F.~F. Damberger, C.~von Schroetter, D.~Theler, P.~Pokorn\'a, F.~H.-T.
  Allain, and J.~{\v S}poner.
\newblock Recognition of {N6}-methyladenosine by the {YTHDC1} {YTH} domain
  studied by molecular dynamics and {NMR} spectroscopy: The role of hydration.
\newblock {\em J. Phys. Chem. B}, 125(28), 7691--7705, (2021).

\bibitem{purslow2021n}
J.~A. Purslow, T.~T. Nguyen, B.~Khatiwada, A.~Singh, and V.~Venditti.
\newblock N 6-methyladenosine binding induces a metal-centered rearrangement
  that activates the human {RNA} demethylase {Alkbh5}.
\newblock {\em Sci. Adv.}, 7(34), eabi8215, (2021).

\bibitem{aduri2007amber}
R.~Aduri, B.~T. Psciuk, P.~Saro, H.~Taniga, H.~B. Schlegel, and J.~SantaLucia.
\newblock {AMBER} force field parameters for the naturally occurring modified
  nucleosides in {RNA}.
\newblock {\em J. Chem. Theory Comput.}, 3(4), 1464--1475, (2007).

\bibitem{stasiewicz2019qrnas}
J.~Stasiewicz, S.~Mukherjee, C.~Nithin, and J.~M. Bujnicki.
\newblock {QRNAS}: software tool for refinement of nucleic acid structures.
\newblock {\em BMC Struct. Biol.}, 19(1), 1--11, (2019).

\bibitem{kierzek2022secondary}
E.~Kierzek, X.~Zhang, R.~M. Watson, S.~D. Kennedy, M.~Szabat, R.~Kierzek, and
  D.~H. Mathews.
\newblock Secondary structure prediction for {RNA} sequences including
  {N6}-methyladenosine.
\newblock {\em Nat. Commun.}, 13(1), 1--10, (2022).

\bibitem{frohlking2020toward}
T.~Fr{\"o}hlking, M.~Bernetti, N.~Calonaci, and G.~Bussi.
\newblock Toward empirical force fields that match experimental observables.
\newblock {\em J. Chem. Phys.}, 152(23), 230902, (2020).

\bibitem{mey2020best}
A.~S. Mey, B.~K. Allen, H.~E.~B. Macdonald, J.~D. Chodera, D.~F. Hahn, M.~Kuhn,
  J.~Michel, D.~L. Mobley, L.~N. Naden, S.~Prasad, et~al.
\newblock Best practices for alchemical free energy calculations [article v1.
  0].
\newblock {\em Living J. Comp. Mol. Sci.}, 2(1), 18378, (2020).

\bibitem{cesari2019fitting}
A.~Cesari, S.~Bottaro, K.~Lindorff-Larsen, P.~Ban{\'a}{\v s}, J.~{\v S}poner,
  and G.~Bussi.
\newblock Fitting corrections to an {RNA} force field using experimental data.
\newblock {\em J. Chem. Theory Comput.}, 15(6), 3425--3431, (2019).

\bibitem{alenaizan2021proto}
A.~Alenaizan, J.~L. Barnett, N.~V. Hud, C.~D. Sherrill, and A.~S. Petrov.
\newblock The proto-nucleic acid builder: a software tool for constructing
  nucleic acid analogs.
\newblock {\em Nucleic Acids Res.}, 49(1), 79--89, (2021).

\bibitem{abraham2015gromacs}
M.~J. Abraham, T.~Murtola, R.~Schulz, S.~P{\'a}ll, J.~C. Smith, B.~Hess, and
  E.~Lindahl.
\newblock {GROMACS}: High performance molecular simulations through multi-level
  parallelism from laptops to supercomputers.
\newblock {\em SoftwareX}, 1, 19--25, (2015).

\bibitem{bernetti2020pressure}
M.~Bernetti and G.~Bussi.
\newblock Pressure control using stochastic cell rescaling.
\newblock {\em J. Chem. Phys.}, 153(11), 114107, (2020).

\bibitem{cornell1995second}
W.~D. Cornell, P.~Cieplak, C.~I. Bayly, I.~R. Gould, K.~M. Merz, D.~M.
  Ferguson, D.~C. Spellmeyer, T.~Fox, J.~W. Caldwell, and P.~A. Kollman.
\newblock A second generation force field for the simulation of proteins,
  nucleic acids, and organic molecules.
\newblock {\em J. Am. Chem. Soc.}, 117(19), 5179--5197, (1995).

\bibitem{perez2007refinement}
A.~P{\'e}rez, I.~March{\'a}n, D.~Svozil, J.~{\v S}poner, T.~E. Cheatham~III,
  C.~A. Laughton, and M.~Orozco.
\newblock Refinement of the {AMBER} force field for nucleic acids: improving
  the description of $\alpha$/$\gamma$ conformers.
\newblock {\em Biophys. J.}, 92(11), 3817--3829, (2007).

\bibitem{zgarbova2011refinement}
M.~Zgarbov\'a, M.~Otyepka, J.~\v{S}poner, A.~Ml\'{a}dek, P.~Ban\'{a}\v{s},
  T.~E. Cheatham, and P.~Jure\v{c}ka.
\newblock Refinement of the cornell et al. nucleic acids force field based on
  reference quantum chemical calculations of glycosidic torsion profiles.
\newblock {\em J. Chem. Theory Comput.}, 7(9), 2886--2902, (2011).

\bibitem{jorgensen1983comparison}
W.~L. Jorgensen, J.~Chandrasekhar, J.~D. Madura, R.~W. Impey, and M.~L. Klein.
\newblock Comparison of simple potential functions for simulating liquid water.
\newblock {\em J. Chem. Phys.}, 79(2), 926--935, (1983).

\bibitem{joung2008determination}
I.~S. Joung and T.~E. Cheatham~III.
\newblock Determination of alkali and halide monovalent ion parameters for use
  in explicitly solvated biomolecular simulations.
\newblock {\em J. Phys. Chem. B}, 112(30), 9020--9041, (2008).

\bibitem{beutler1994avoiding}
T.~C. Beutler, A.~E. Mark, R.~C. van Schaik, P.~R. Gerber, and W.~F.
  Van~Gunsteren.
\newblock Avoiding singularities and numerical instabilities in free energy
  calculations based on molecular simulations.
\newblock {\em Chem. Phys. Lett.}, 222(6), 529--539, (1994).

\bibitem{goga2012efficient}
N.~Goga, A.~Rzepiela, A.~De~Vries, S.~Marrink, and H.~Berendsen.
\newblock Efficient algorithms for {Langevin} and {DPD} dynamics.
\newblock {\em J. Chem. Theory Comput.}, 8(10), 3637--3649, (2012).

\bibitem{parrinello1981polymorphic}
M.~Parrinello and A.~Rahman.
\newblock Polymorphic transitions in single crystals: A new molecular dynamics
  method.
\newblock {\em J. Appl. Phys.}, 52(12), 7182--7190, (1981).

\bibitem{darden1993particle}
T.~Darden, D.~York, and L.~Pedersen.
\newblock Particle mesh {Ewald}: An {N} log ({N}) method for {Ewald} sums in
  large systems.
\newblock {\em J. Chem. Phys.}, 98(12), 10089--10092, (1993).

\bibitem{souaille2001extension}
M.~Souaille and B.~Roux.
\newblock Extension to the weighted histogram analysis method: combining
  umbrella sampling with free energy calculations.
\newblock {\em Comput. Phys. Commun.}, 135(1), 40--57, (2001).

\bibitem{shirts2008statistically}
M.~R. Shirts and J.~D. Chodera.
\newblock Statistically optimal analysis of samples from multiple equilibrium
  states.
\newblock {\em J. Chem. Phys.}, 129(12), 124105, (2008).

\bibitem{tan2012theory}
Z.~Tan, E.~Gallicchio, M.~Lapelosa, and R.~M. Levy.
\newblock Theory of binless multi-state free energy estimation with
  applications to protein-ligand binding.
\newblock {\em J. Chem. Phys.}, 136(14), 144102, (2012).

\bibitem{efron1986bootstrap}
B.~Efron and R.~Tibshirani.
\newblock Bootstrap methods for standard errors, confidence intervals, and
  other measures of statistical accuracy.
\newblock {\em Stat. Sci.}, pages 54--75, (1986).

\bibitem{bennett1976}
C.~H. Bennett.
\newblock Efficient estimation of free energy differences from {Monte} {Carlo}
  data.
\newblock {\em J. Comp. Phys.}, 22(2), 245--268, (1976).

\bibitem{wu2005bar}
D.~Wu and A.~D. Kofke.
\newblock Phase-space overlap measures. ii. design and implementation of
  staging methods for free-energy calculations.
\newblock {\em J. Chem. Phys.}, 123(8), 084109, (2005).

\bibitem{zhu1997algorithm}
C.~Zhu, R.~H. Byrd, P.~Lu, and J.~Nocedal.
\newblock Algorithm 778: {L-BFGS-B}: Fortran subroutines for large-scale
  bound-constrained optimization.
\newblock {\em ACM Trans. Math. Softw.}, 23(4), 550--560, (1997).

\bibitem{virtanen2020scipy}
P.~Virtanen, R.~Gommers, T.~E. Oliphant, M.~Haberland, T.~Reddy, D.~Cournapeau,
  E.~Burovski, P.~Peterson, W.~Weckesser, J.~Bright, et~al.
\newblock Scipy 1.0: fundamental algorithms for scientific computing in python.
\newblock {\em Nat. Methods}, 17(3), 261--272, (2020).

\bibitem{GrayKish1969}
P.~G. Gray and L.~Kish.
\newblock Survey sampling.
\newblock {\em J. Royal Stat. Soc. A (General)}, 132(2), 272, (1969).

\bibitem{rangan2018determination}
R.~Rangan, M.~Bonomi, G.~T. Heller, A.~Cesari, G.~Bussi, and M.~Vendruscolo.
\newblock Determination of structural ensembles of proteins: restraining vs
  reweighting.
\newblock {\em J. Chem. Theory Comput.}, 14(12), 6632--6641, (2018).

\bibitem{cieplak1995application}
P.~Cieplak, W.~D. Cornell, C.~Bayly, and P.~A. Kollman.
\newblock Application of the multimolecule and multiconformational {RESP}
  methodology to biopolymers: Charge derivation for {DNA}, {RNA}, and proteins.
\newblock {\em J. Comput. Chem.}, 16(11), 1357--1377, (1995).

\bibitem{norgaard2008experimental}
A.~B. Norgaard, J.~Ferkinghoff-Borg, and K.~Lindorff-Larsen.
\newblock Experimental parameterization of an energy function for the
  simulation of unfolded proteins.
\newblock {\em Biophys. J.}, 94(1), 182--192, (2008).

\bibitem{li2011iterative}
D.-W. Li and R.~Br{\"u}schweiler.
\newblock Iterative optimization of molecular mechanics force fields from {NMR}
  data of full-length proteins.
\newblock {\em J. Chem. Theory Comput.}, 7(6), 1773--1782, (2011).

\bibitem{schroeder2009optical}
S.~J. Schroeder and D.~H. Turner.
\newblock Optical melting measurements of nucleic acid thermodynamics.
\newblock In {\em Meth. Enzymol.}, volume 468, pages 371--387. Elsevier,
  (2009).

\bibitem{janecek2021w}
M.~Jane{\v{c}}ek, P.~K{\"u}hrov{\'a}, V.~Ml{\`y}nsk{\`y}, M.~Otyepka, J.~{\v
  S}poner, and P.~Ban\'{a}\v{s}.
\newblock {W-RESP}: Well-restrained electrostatic potential-derived charges.
  revisiting the charge derivation model.
\newblock {\em J. Chem. Theory Comput.}, 17(6), 3495--3509, (2021).

\bibitem{singh1984approach}
U.~C. Singh and P.~A. Kollman.
\newblock An approach to computing electrostatic charges for molecules.
\newblock {\em J. Comput. Chem.}, 5(2), 129--145, (1984).

\bibitem{Kameda2015WCenergies}
S.~Sakuraba, K.~Asai, and T.~Kameda.
\newblock Predicting {RNA} duplex dimerization free-energy changes upon
  mutations using molecular dynamics simulations.
\newblock {\em J. Phys. Chem. Lett.}, 6(21), 4348--4351, (2015).

\bibitem{condon2015stacking}
D.~E. Condon, S.~D. Kennedy, B.~C. Mort, R.~Kierzek, I.~Yildirim, and D.~H.
  Turner.
\newblock Stacking in {RNA}: {NMR} of four tetramers benchmark molecular
  dynamics.
\newblock {\em J. Chem. Theory Comput.}, 11(6), 2729--2742, (2015).

\bibitem{bottaro2018conformational}
S.~Bottaro, G.~Bussi, S.~D. Kennedy, D.~H. Turner, and K.~Lindorff-Larsen.
\newblock Conformational ensembles of {RNA} oligonucleotides from integrating
  {NMR} and molecular simulations.
\newblock {\em Sci. Adv.}, 4(5), eaar8521, (2018).

\bibitem{kuhrova2019improving}
P.~K{\"u}hrov{\'a}, V.~Ml{\`y}nsk{\`y}, M.~Zgarbov{\'a}, M.~Krepl, G.~Bussi,
  R.~B. Best, M.~Otyepka, J.~{\v S}poner, and P.~Ban\'{a}\v{s}.
\newblock Improving the performance of the {AMBER} {RNA} force field by tuning
  the hydrogen-bonding interactions.
\newblock {\em J. Chem. Theory Comput.}, 15(5), 3288--3305, (2019).

\bibitem{guckian2000factors}
K.~M. Guckian, B.~A. Schweitzer, R.~X.-F. Ren, C.~J. Sheils, D.~C. Tahmassebi,
  and E.~T. Kool.
\newblock Factors contributing to aromatic stacking in water: evaluation in the
  context of {DNA}.
\newblock {\em J. Am. Chem. Soc.}, 122(10), 2213--2222, (2000).

\bibitem{isaksson2005uniform}
J.~Isaksson and J.~Chattopadhyaya.
\newblock A uniform mechanism correlating dangling-end stabilization and
  stacking geometry.
\newblock {\em Biochemistry}, 44(14), 5390--5401, (2005).

\end{thebibliography}

\end{document}